\documentclass[article,twocolumn,showpacs,superscriptaddress,amsmath,amssymb]{revtex4}
\usepackage{cancel}
\usepackage{maybemath}
\usepackage{xcolor}
\usepackage{graphicx}

\begin{document}
Published as a final section of:\\
R. Ehrlich, Lett. in High En. Phys. 199, 2021.\\

\title{The KATRIN neutrino mass results: An alternative interpretation}
\author{Robert Ehrlich}
\affiliation{George Mason University, Fairfax, VA 22030}
\email{rehrlich@gmu.edu}
\date{\today}

\begin{abstract}
The KATRIN experiment has recently reported a new upper limit to the mass of the electron neutrino of $m<0.8 eV$ (90\%CL), and a best value of $m^2= 0.26 \pm 0.34 eV^2/c^4$ based on a fit to the observed tritium beta decay spectrum.  Here another interpretation of their results is discussed.\\

keywords: neutrino mass
\end{abstract}
\maketitle

\section{Introduction}
The latest KATRIN result,\citep{Ak2021} the first to achieve sub-eV sensitivity in the neutrino mass, was based on their first and second data-taking campaigns, and it comes two years after an initial result from 2019.\citep{Ak2019}  Those initial results were based on a fit to the experimental beta spectrum recorded at 27 set point energies, using these four free parameters: the signal amplitude $A_s,$ the effective $\beta-$decay end point $E_0,$ the background rate $R_{bg}$, and the neutrino mass square $m^2.$  Four free parameters had then been considered the appropriate choice for doing a ``shape-only” fit.  In contrast to that initial result, the result reported in May 2021 fit the observed spectrum using 37 free parameters.  The implications of employing so many parameters, and the possible motivation for doing so are explored in this short paper.

\section {Fitting the beta spectrum}

KATRIN and virtually all previous neutrino mass experiments have fit their observed spectra using a single effective electron neutrino mass, a choice justified by the small separation between the masses contributing to the electron neutrino, as stipulated in the standard neutrino mass model.  In 2013, the author introduced a highly speculative $3+3$ neutrino mass model derived from an analysis of SN 1987A data.  The model postulated three active-sterile doublets having three specific masses: $m_1=4.0\pm 0.5 eV,$ $m_2=21.4 \pm 1.2 eV,$ and $m_3^2\sim -0.2 keV^2.$~\citep{Eh2013}  Although the model is highly exotic, and involves a ``tachyonic" mass state, it has satisfied a number of empirical tests.~\citep{Eh2018}  Note that the tachyonic doublet here has nothing to do with the negative $m^2_\nu$ reported in some past neutrino mass experiments, which had another cause,~\citep{Be1995} and certainly was not indicative of a tachyonic neutrino.

One of the empirical tests supporting the model involved fits to three neutrino mass experiments prior to KATRIN, where it was claimed that the $3+3$ model fit the data better than the conventional one of a single effective mass.~\citep{Eh2016}  Regrettably, that paper was in error, since it relied on an anomaly seen in the data at $E-E_0\approx - 20 eV$ as evidence for the presence of a mass $m_2\approx 20eV,$ and it therefore assumed too large a contribution to the spectrum from that mass.   In fact, that anomaly near $-20$ eV was mainly due to final state distributions, and those pre-KATRIN experiments were simply not sensitive enough to properly test the $3+3$ model, but they also did not contradict it.  

To fit a beta spectrum using such unequal masses as in the $3+3$ model one  forms an incoherent sum of the spectra for each mass using appropriately chosen weighting factors.  Independent of the $3+3$ model evidence has been presented that the electron neutrino effective mass squared, defined in the usual way: $m_{\rm{i,eff}}^2=\Sigma{|U_{ij}|^2m^2_j},$ has the value $- 0.11 eV^2.$\citep{Eh2015}  With this constraint, if one of the three spectral contributions is stipulated, say that for the 4.0 eV mass, the other two immediately follow.

Following publication of KATRIN's first results in 2019 the author showed that they were consistent with his $3+3$ model as well as the standard single effective mass model, but only for a narrow range of contributions for the 4.0 eV mass, specifically for $\alpha=|U_{i,1}|^2=0.94\pm 0.02.$\citep{Eh2019}  The need for a narrow range in $\alpha$ is illustrated in Fig. 1, which shows the difference in the $\emph {integral}$ spectra for the two models, computed based on phase space.  Note that the same phase space function has been used for the $m^2>0$ and $m^2<0$ neutrino masses, and the spectra for both are assumed to vanish for $E-E_0>0.$
\begin{figure}
\centerline{\includegraphics[angle=0,width=1.0\linewidth]{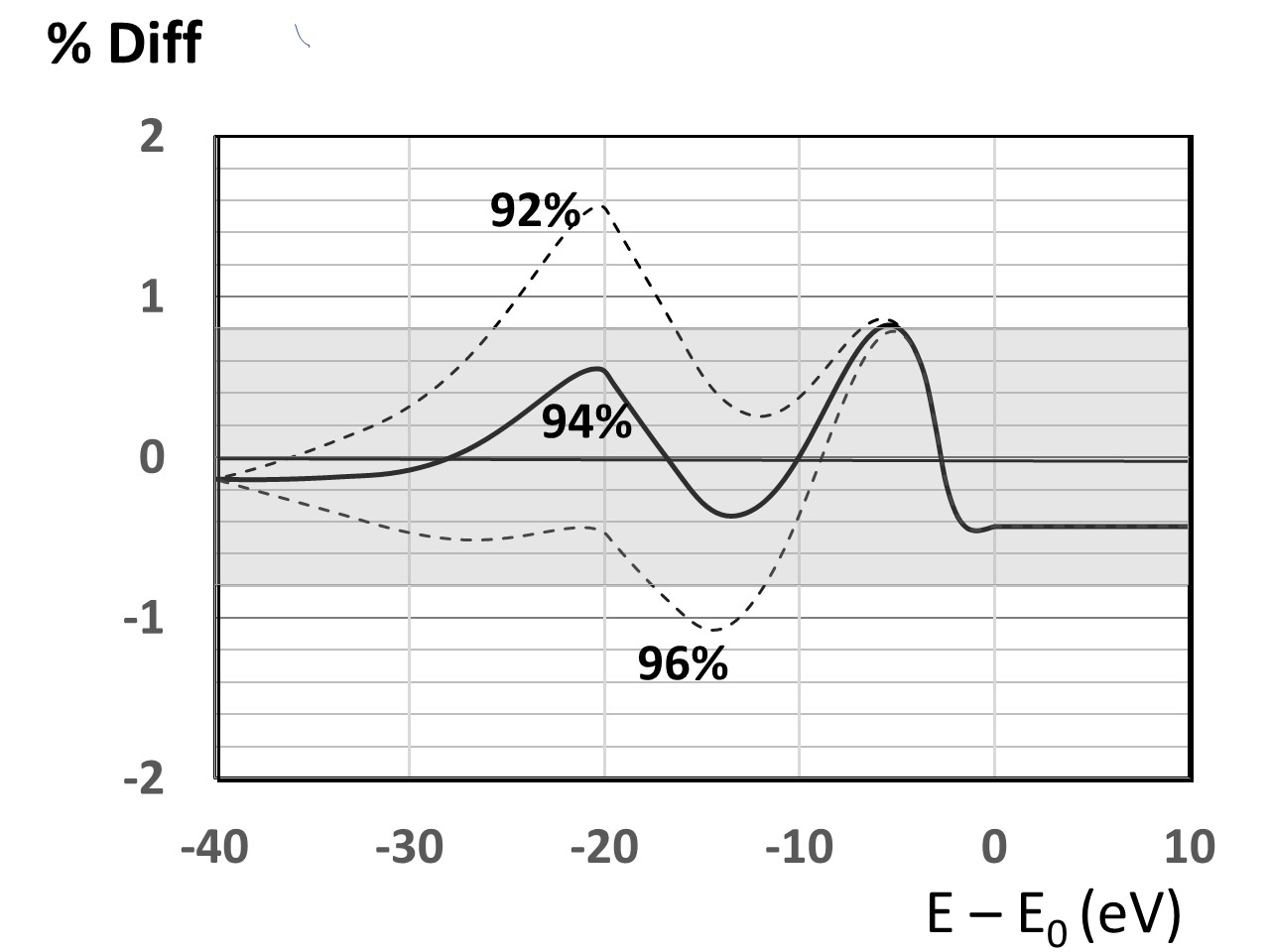}}
\caption{\label{fig1}Computed difference between the integral spectra for the $3+3$ model of the neutrino masses and the standard model of a small single effective mass.  For the $3+3$ model three values of the contribution for the 4.0 eV mass are shown: $92\%, 94\%$ and $96\%.$  Only for the $94\%$ case are the two integral spectra everywhere within $\pm 0.8\%$ of each other.  The graph shows the difference in the two integral spectra, because the integral spectrum is what KATRIN measures.}
\end{figure}

A consequence of Fig. 1 is that if the KATRIN data are well described by the $3+3$ model, but they are fit to a single small effective mass spectrum, then one would expect to see two ``bumps," that is, excess 
numbers of events ($O(0.5\%)$) at a distance from $E_0$ close to the two $m^2>0$ masses in the model.  The initial KATRIN results from 2019 were such that the statistical error bars were not small enough to clearly show whether these excesses were present, although the residuals for their best fit did show an indication they were.  In fact, the ninth residual in their fit located at $E-E_0 = -23 eV$, which was consistent with the position of the left peak in Fig. 3, fell $+2.5\sigma$ above their best four free parameter fit to a single effective mass.\citep{Eh2019}.  A similar hint of an excess was found in KATRIN's second run in 2021.  However, that most recent KATRIN experiment fit to the spectrum released in May 2021 has used not four free parameters to fit the data, but rather 37.~\citep{Ak2021}.  One of those parameters is the value of $m^2$ and the other 36 are the signal amplitude, $S_i$, background amplitude, $B_i,$ and endpoint energy, $E_{0,i},$ for the data recorded in each of 12 concentric rings on the detector.  While the assumption of a radial dependence of signal and background is reasonable in one sense, it was not done previously.  Moreover, the use of so many free parameters can mask the $O(0.5\%)$ departures from the single mass spectrum that would occur if the $3+3$ model were a valid description of the data -- see Fig. 1.  In fact, the number of adjustable parameters KATRIN uses in its fits is even greater than 37 if one counts some unspecified number of ``pull terms" corresponding to constrained parameters that can vary within limits.

\section {Fitting (or hiding) an elephant}
The mathematician John Von Neumann famously once said: ``With four parameters I can fit an elephant, and with five I can make him wiggle his trunk."~\citep{Dy2004}  This quote was his humorous way of telling us to be suspicious of using too many free parameters in doing a fit.  Despite Von Neumann's caution, the use of four parameters by KATRIN to fit its initial data was exactly appropriate, but 37-plus parameters it now uses would seem to be excessive.   Whether the departures from the single effective mass spectrum indicated in Fig. 3 are in fact present in the KATRIN data remains uncertain at present, and will probably remain so, unless the number of free and partly free parameters used to fit the data is significantly reduced.  Moreover, given that hints of a good fit to the $3+3$ spectrum have now appeared in each of KATRIN's first two runs, one would hope that they find a way to combine the results of all their runs into a single fit so as to reveal any small departures from the single mass spectrum, as indicated in Fig. 1.


\begin{thebibliography}{10}
\bibitem{Ak2021}M. Aker et al. (KATRIN Collaboration);\\arXiv:2105.08533.
\bibitem{Ak2019}M. Aker et al. (KATRIN Collaboration), Phys. Rev. Lett. {\bf123}, 221802 (2019).
\bibitem{Eh2013} R. Ehrlich, Astropart. Phys., {\bf41} 1-6, (2013);\\arxiv.org/1204.0484.
\bibitem{Eh2018} R. Ehrlich, Adv. in Astron., 2820492 (2019);\\arXiv.org/1711.09897.
\bibitem{Be1995} A. I. Belesev at al., Phys. Lett. B, {\bf350}, (2), 263-272 (1995).
\bibitem{Eh2016} R. Ehrlich, Astropart. Phys., {\bf85}, 43-49 (2016);\\
arxiv.org/1602.09043
\bibitem{Eh2015} R. Ehrlich, Astropart. Phys. {\bf66}, 11 (2015);\\
arxiv.org/1408.2804
\bibitem{Eh2019} R. Ehrlich, Lett. in High En. Phys., {\bf4}, (4), (2019);\\
http://journals.andromedapublisher.com/\\
index.php/LHEP/article/view/139/76
\bibitem{Dy2004} F. Dyson, Nature, 427, 6972, 297 (2004).
\end{thebibliography}
\end{document}